\documentclass[a4paper,10pt]{article}
\usepackage[utf8]{inputenc}
\usepackage[dvips]{color}
\usepackage{algorithm}
\usepackage{algorithmic}
\usepackage{enumitem}
\usepackage[english]{babel}
\usepackage[numbers]{natbib}
\usepackage{tikz}
\title{diagno-syst: a tool for accurate inventories in metabarcoding}
\author{J.-M. Frigerio$^{1,2}$, F. Rimet$^3$,  A. Bouchez$^3$, \\
E. Chancerel$^1$, P. Chaumeil$^{1,2}$, F. Salin$^{1,2}$, S. Th\'erond$^4$,\\
M. Kahlert$^5$ \& A. Franc$^{1,2*}$\\
\footnotesize{$^1$ INRA, UMR BioGeCo, Bordeaux, France}\\
\footnotesize{$^2$ INRIA, Pleiade team, Bordeaux, France}\\
\footnotesize{$^3$ INRA, UMR Carrtel, Thonon-les-Bains, France}\\
\footnotesize{$^4$ CNRS, IDRIS, Orsay, France}\\
\footnotesize{$^5$ SLU, Uppsala, Sweden}\\
\footnotesize{$^*$ to whom correspondence should be addressed:\texttt{alain.franc@inra.fr}}
}

\begin{document}

\maketitle




\begin{abstract}
Metabarcoding on amplicons is rapidly expanding as a method to produce molecular based inventories of microbial communities. Here, we work on 
freshwater diatoms, which are microalgae possibly inventoried both on a morphological and a molecular
basis. We have developed an algorithm, in a program called \texttt{diagno-syst}, based a the notion of informative read, which carries out 
supervised clustering of reads by mapping them exactly one by one on all reads of a well curated and taxonomically annotated reference database. This program has been 
run on a HPC (and HTC) infrastructure to address computation load. We 
compare optical and molecular based inventories on 10 samples from L\'eman lake, and 30 from Swedish rivers. We track
all possibilities of mismatches between both approaches, and compare the results with standard pipelines (with heuristics) like Mothur. We find that
the comparison with optics is more accurate when using exact calculations, at the price of a heavier computation load. It is crucial when 
studying the long tail of biodiversity, which may be overestimated by pipelines or algorithms using heuristics instead (more false positive). 
This work supports the analysis that these methods will benefit from progress in, first, building an 
agreement between molecular based and morphological based systematics and, second, having as complete as possible publicly available reference 
databases. 
\end{abstract}

\section{Introduction}

Unexpected diversity of unicellular Eucaryots (nanoplankton) or hydrothermal sediments has been revealed in 2001 
by sequencing
ribosomal DNA (18S), using BLAST for comparison with databases, and establishing molecular phylogenies of
these unexplored worlds, \cite{Lopez-Garcia2001,Lopez-Garcia2003}. At the same time, unexpected diversity of oceanic picoplankton 
has been revealed using 18S and phylogenies too \cite{Moon2001}. Since then, exploration of microbial diversity,
be it in protists, bacteria or archaea by molecular based tools has exploded, and has become a standard for 
biodiversity assessments. Soon after, Hebert and coll. launched the barcoding of life for animals \cite{hebert03},
which has been quickly extended to other kingdoms (see e.g. \cite{Hollingsworth2009}), and to protists \cite{Pawlowski2012}. 
The advance of sequencing technologies, especially NGS, producing
first hundreds of thousands of reads with pyrosequencing, then millions with Ilumina, have paved the way for metagenomics, 
and metabarcoding for biodiversity or biomonitoring studies (see e.g., among many others,  
\cite{Hajibabaei2011, Bik2012,Taberlet2012,Kermarrec2014,Dafforn2014,Joly2014,Pawlowski2014}).\\
\\
Metabarcoding is expected to yield inventories of species diversity similar to
those provided by morphological based methods, if molecular and morphological taxonomies agree. One way for a rigorous 
verification is by isolating and culturing, and 
comparing
with results of metabarcoding. \cite{Pedros-Alio2007} cites a well known case of a bacteria in 
the Mediterranean Sea (\emph{Leeuwenhoekiella blandensis}), which has been isolated and cultivated but never found in
repeated molecular inventories on the same spot. A still more controlled way to estimate the quality of the information given by 
metabarcoding is to build artificial samples in 
the laboratory, and run the metabarcoding protocol on these laboratory-assembled communities (see 
\cite{Porazinska2009} for nematodes, \cite{Kermarrec2013} for diatoms), or in silico experiments (see e.g. \cite{Clarke2014}). \\
\\
A second way to estimate this quality, on which we will focus here, is by comparing morphological based inventories with 
molecular based inventories on
same natural communities. Such an approach has been used for alveolates in a freshwater lake
 \cite{Medinger2010}, tintinnids (ciliates, microzooplankton) \cite{Bachy2013}, arthropods and birds \cite{Ji2013}, seegrass 
 communities \cite{Cowart2015}, 
 freshwater diatoms \cite{Zimmerman2015}, the latter insisting on the need to derive tools for assigning a taxon name to a sequence 
 (what  will be referred to in the sequel as taxonomic assignment), and estuarine plankton \cite{Abad2016}, where completing the 
 reference database  was found to settle most of inconsistencies between both approaches in zooplankton. Such an approach permits to study 
 a still open question: 
 how reliable is metabarcoding for inventorying or unveiling rare species? How to estimate the so called \emph{rare biosphere} 
 \cite{Sogin2006,Pedros-Alio2012,Lindeque2013,Debroas2015} ? But see \cite{Reeder2009,Kunin2010}.\\
\\
We develop here a data analysis framework without any heuristics, doing exact computations using High Performance Computing
techniques, in order to produce as accurate as possible molecular based inventories, quantify the comparison between morphological based and 
molecular based inventories, facilitate 
the localization of major sources of errors, and propose some priorities to fix them.\\
\\
Discrepancies in comparing a morphological based and a molecular based inventory on the same sample can have 
three origins: $(i)$ inappropriate morphological based inventory $(ii)$ inappropriate molecular based inventory $(iii)$ both
inventories are correct but do not give the same information (differences in species delineation between molecular phylogenies 
and 
morphological based identification, for example). The latter is crucial, and deserves further research. As an example, 
accurate comparison between molecular and morphological taxa delineation in angiosperms has been shaped over many decades 
through collaboration between many teams (see \cite{APGIII} in 2009, after APG in 1998 and APGII in 2003). Such a comparison remains crucial, as 
a benchmark for ongoing 
characterization 
of currently unseen biodiversity in eukaryotic communities. Here, we develop a tool, named \verb%diagno-syst%, for having an as accurate as 
possible molecular 
based inventory, relying on High Performance Computing for handling massive data, without any heuristics to speed up calculation process. We 
focus on best possible quality for molecular based inventories, assuming that morphological based inventories are correctly evaluated (i.e. we will 
not discuss here morphological based taxonomy of diatoms).\\
\\
One of the requirements for metabarcoding is selecting markers which are, 
as much as possible, universal, resolutive, and 
technologically easy to work with. Most of this
work is inherited from barcoding. A synthesis paper on marker choice for Eukaryotes from CBOL Protist Working Group is 
\cite{Pawlowski2012}. In the remaining of this paper, we assume that a 
marker has been selected. Knowing that, several sources of errors have been recognized in 
metabarcoding data analysis (see \cite{Bachy2013} and references therein). There exists several studies 
or reviews 
(see e.g. \cite{Zinger2012,Bucklin2016}) which focus on well known pitfalls and caveats of molecular approaches, but very few on the 
pitfalls and caveats when using heuristics instead of exact computations because of the size of data sets. We focus here 
on isolating bioinformatics challenges from biological challenges, by running exact calculation for sequence comparisons, without any
heuristics.\\
\\
A molecular based inventory is built on sequence comparisons between a set of queries and a reference database, often on 
amplicons. Such a comparison can be made exactly by Needleman-Wunsch algorithm 
\cite{NW70} for
global alignment or Smith-Waterman algorithm \cite{SW81} for local alignment. These algorithms are well understood and exact
\cite{gusfield97}.
They work well on small size data sets as those produced by Sanger technologies. However, they are quadratic in time as scaling with 
the number of queries $\times$ the number of references. BLAST \cite{Altschul1990} is a well established tool for exploring
quickly a huge database of references. We have the opposite in metabarcoding communities of a given group of organisms: the 
reference database is often of reasonable size (a few
thousands specimen), whereas the number of queries to match becomes huge (about $10^5$ or even $10^6$). Moreover, it is known that 
BLAST can produce bad results \cite{Guillou2013}. Therefore, as in \cite{Guillou2013}, we developed specific assignation tools
based on local alignment (here, Smith-Waterman score). As being exact is paid by a heavier computation load, we have exploited the fact that 
this algorithm belongs to the category of massively parallel problem: the $10^5$ or $10^6$ local alignments of each query with all specimen 
in the 
reference database are all independent, and can be run in parallel on different cores of a computer. Hence, we have opted
for a massively parallel implementation of exact local alignment, which answers exactly to the raised question of sequence comparison between 
a reference database and a set of queries.\\
\\

\section{Material}

We have worked on inventories of diatom communities, as diatom species can be identified both optically (by looking at the frustule), 
and molecularly (many reference sequences for diatom barcodes are available). This enables to test the quality of the dictionary between 
both approaches.\\
\\
The starting point for data analysis is a set of several sets: each sample is given as a set of sequences (the reads), and we have a set of samples. 
We have one reference database, R-Syst::Diatoms \cite{Rimet2016}, which will 
be used for all environmental
samples. The reference database is given as a set of $n$ specimens. For 
each specimen, we have a sequence identifier (a code), a sequence (of the same marker with the same primers), and a hierarchical list
of names, naming Order, Family, Genus and Species. We have a set of 10 environmental samples from Léman lake, and a second set of 
20 environmental samples from Swedish rivers. 
For each environmental sample, we have an optic inventory, made by a trained diatomist, and a set of reads, which are produced after
amplification of a given marker with given primers. An optical inventory has been produced for each environmental sample by counting 
about 400 diatom valves in a microscope and identifying them via standard diatom literature using CEN (European Standardization Committee) 
methods \cite{AFNOR2014,AFNOR2016}.

\subsection{Environmental samples}

Two datasets were analyzed: one from Léman lake, and one from Sweden. The Léman lake dataset contains one site sampled at ten different dates
(about one month between two sampling campaigns). The Swedish dataset contains data of 13 streams and one lake distributed over the whole 
country, six of the streams sampled in two subsequent years. The streams covered different ecological regions and stream types, including 
lowland agricultural streams common also in Central Europe, small boreal streams and mountain streams with no forest cover in the catchment.
In this way, we could test if the reference data base of diatoms developed to match Léman Lake communities can be used also for other regions 
of the 
world. We expected more problems when using it for Swedish data.

\subsection{Reference database: names}

The taxonomic names selected in this study are those selected in Thonon Reference Database, \texttt{R-Syst::Diatoms}, which has 
been built first at
Thonon (Thonon Culture Collection, TCC, see \texttt{www.inra.fr/carrtel-collection}), and complemented by careful curation
of freshwater diatom sequences in NCBI \cite{Kermarrec2013}. The privileged taxonomic level is the species level. Names in the 
reference database are as much as possible accepted authored names. However, 
in several circumstances, a discrepancy between molecular based species delineation by phylogenies and morphological based
species delineation have been acknowledged. In such a case, a compromise has been made (e.g. \emph{Fragilaria group}) in order
to have the best agreement between morphological based and molecular based taxonomic identification of specimen in the 
reference database. It is not the purpose of this paper to discuss diatom systematics, and the current state of the art, even
if acknowledged as perfectible, is taken for granted and accepted. Any accepted evolution or change in diatom's systematics 
can be taken into account in a 
further step just by changing the accepted names. Taxa in this study should be considered as operational taxonomic units as
close as possible to species level and accepted species names, knowing that some discrepancies still exist in the current 
state of the art. For sake of simplicity, they will be referred to as \emph{species} in the sequel of this paper.

\subsection{Reference database: sequences}

\paragraph{Marker choice:} Before the 
bloom of barcoding, molecular taxonomy of plankton has been studied using 18S as a marker. Over the subsequent years, several 
hypervariable regions of 18S 
have been progressively selected in scientific community to study plankton, and more generally protists, i.e. unicellular eukaryots 
(see \cite{Pawlowski2012,Bucklin2016,Tanabe2016}). A second criteria for a marker choice is the availability of a well curated database.
PR$^2$ (see \cite{Guillou2013} and \texttt{http://ssu-rrna.org/}) has been designed specifically for marine plankton. \cite{Kermarrec2013} 
have compared rbcL, 18S and CO1 for freshwater diatoms, the three of them being
acknowledged as potential barcodes for diatoms, making a balance between resolution and richness of existing database. They have shown that
rbcL was more resolutive than 18S, and that public information was available for both of them. Hence, a choice has been made to work with 
rbcL, and to complete the existing database. Subsequently, an inner marker has been selected, of 312 bp long to comply with technological 
requirements of NGS sequencing facilities, which could not provide $\geq 500 bp$ long reads as Sanger technology could. For amplifying this 
region, the primer pair \verb%Diat_rbcL_708F% \cite{Stoof-Leichsenring2012} and \verb%R3% \cite{Bruder2007} 
was used directly for samples from Léman lake, and was modified for the Swedish samples
as follow: forward primer combine an equimolar mix of \verb%Diat_rbcL_708F_1% (AGGTGAAGTAAAAGGTTCWTACTTAAA), \verb%Diat_rbcL_708F_2% 
(AGGTGAAGTTAAAGGTTCWTAYTTAAA) and \verb%Diat_rbcL_708F_3% (AGGTGAAACTAAAGGTTCWTACTTAAA); reverse primer combine an equimolar mix of 
\verb%R3_1% (CCTTCTAATTTACCWACWACTG) and \verb%R3_2% (CCTTCTAATTTACCWACAACAG). The resolution of the selected short barcode has been checked, and 
the drop from full rbcL fragment, even if noticeable, was found to be non significant. 

\section{Methods}

For each set of samples (Léman lake and Swedish rivers), the work has been processed along a sequence of three steps
\begin{itemize}
 \item have an optical based inventory
 \item build a molecular based inventory
 \item identify and classify mismatches between both.
\end{itemize}
Methods for each of these steps are given in what follows, with a focus on molecular based inventories and classification of mismatches.\\
\\
Let us note that an output of a Proton sequencer for an environmental sample typically counts about $10^5$ reads, of $312 bp$ long each. Here, 
instead of using heuristics to speed up the analysis, we have kept on the choice to use exact algorithm, which scale linearly with the number
of reads. We have designed as well automatic procedures to handle a large number of environmental samples, i.e. data flows between different steps
in calculation. Therefore, we have used a High Performance Computing Center for intensive computing (IDRIS, see \texttt{http://www.idris.fr/eng/}),
and iRODS as a file manager for managing data flows.

\subsection{Molecular based inventories: data analysis}

The algorithms used for taxonomic annotation of queries knowing a reference database (supervised clustering) are given here. The motivations for 
such a choice, i.e. the notion of informative read and sliding barcoding gap, which distinguish these algorithms from, say, those encapsulated in
Mothur, are presented in the discussion section (section \ref{subsec:inventories}).

\paragraph{Notations:} Let us give some notations. The set of references is denoted $\mathbf{R}$. Each reference is a 312bp long sequence,
with little variations in length, with an identifier, and a taxonomic annotation (species, genus, family). There is a set
of queries per environmental sample $i$, denoted $\mathbf{Q}_i$. Each query is a 312bp long sequence, with some variatons
in length, with an identifier. The number of references is denoted $|\mathbf{R}|$, and the number of queries
in sample $i$ is denoted $|\mathbf{Q}_i|$. We have for this study $|\mathbf{R}|=1446$. The barcode gap is denoted $\alpha$, and maximum barcode gap 
$\theta$.

\paragraph{Distance between sequences:} We have implemented in C the Smith-Waterman algorithm \cite{SW81} producing a score of 
local alignment between 
two sequences. Let $(q,r)$ be a pair query $\times$ reference. Let $sw(q,r)$ 
be the Smith-Waterman score between the query and the 
reference. Then $sw(q,r)$ is the highest sore of global alignment between
a substring $\alpha \subset q$ and a substring $\beta \subset r$ over all pairs $(\alpha,\beta)$ (see \cite{gusfield97}). 
Selected 
costs are 
$+1$ for a match, and $-1$ for a gap (for one base) or a substitution. This can easily be extended to more diverse evolution
models. The score has been 
translated into a distance as follows: let $\ell(q)$ be the length of the query, $\ell(r)$ be the length of the reference. Then
\begin{equation}
 d(q,r) = \frac{\min \{\ell(r), \ell(q)\}- sw(q,r)}{2}
\end{equation}
The program computing $d(q,r)$ from $q$ and $r$ as inputs is called \texttt{disseq}. 

\paragraph{Step 1:} Step 1 is to compute $d(q,r)$ for any pair $(q,r) \in \mathbf{Q} \times \mathbf{R}$. This has been done 
by a double loop. This is 
the intensive part of the computation. It has been run both on Babel and Turing. Babel is a BlueGene P (IBM) located at Idris
in Saclay. It offers massive parallelization, fully adapted to the computation of large matrices of pairwise distances. Its 
characteristics are:  40,960 cores PowerPC 450, 20 To of memory, 139 Tflop/s peak power, 
and 800 To of memory on disks.  Turing is a BlueGene Q (IBM) located at IDRIS too. Its architecture is made of 6 racks,
of 1024 nodes each, and 16 cores per node, hence 98 304 cores. We have used Turing with up to $2^{14}= 16384$ cores (one rack).
Its peak power is 1,258 Pflop/s. Their main advantage (apart from massive parallelization) is a low energy consumption: 300 kW,
at a cost of lower frequency in cores, but larger number of cores. Such an architecture is particularly suitable for massive embarassingly
parallel jobs. The loop for computing all pairwise distances
has been run with Message Passing Interface (MPI) programming, by a dedicated program called \texttt{mpi-disseq}, written 
in C and MPI for this purpose. The program has been tested first on Avakas (M\'esocentre de Calcul Intensif Aquitain, 264
computing nodes, 12 cotes each) and then ported to Babel and Turing. The program scales perfectly.

\paragraph{Step 2:} A barcoding gap $\alpha$ being selected, step 2 is to derive the neighborhood of any query $q \in\mathbf{Q}$ 
as the set $\mathcal{N}_\alpha(q)$ of all sequences in 
$\mathbf{R}$ such that $d(q,r) \leq \alpha$
\begin{equation}
  \mathcal{N}_{\alpha}(q) = \{r \in \mathbf{R} \: : \: d(q,r) \leq \alpha\}
\end{equation}

\paragraph{Step 3:} This step is assigning a name, or not, to a query, knowing its neighborhood in $\mathbf{R}$. A query
$q$ is called \emph{informative} if all the references $r \in \mathcal{N}_{\alpha}(q)$ belong to the same taxon. A query can be
informative for the genus, but not for the species. Hence, a small program has been written (in python) which implements the 
following procedure:\\
\\
\begin{algorithm}[H]
\begin{algorithmic}[1]
\STATE \textbf{input:} $q \in \mathbf{Q}\qquad$ 
\COMMENT{query}
\STATE \textbf{input}  $\alpha\qquad\qquad$
\COMMENT {barcoding gap}
\STATE get $\mathcal{N}_{\alpha}(q)$
\IF{$\mathcal{N}_{\alpha}(q) == \emptyset$}
\STATE \texttt{tax\_annot} $\leftarrow$ \texttt{'unknown'}
\ELSE
\STATE $T \: \leftarrow \: $ list of taxa in $\mathcal{N}_{\alpha}(q)$ 
\STATE $|T| \: \leftarrow \:$ number of taxa in $T$
\IF{$|T|==1$}
\STATE \texttt{tax\_annot} $\leftarrow$ the unique taxon in $T$
\ELSE
\STATE \texttt{tax\_annot} $\leftarrow$ \texttt{'ambiguous'}
\ENDIF
\ENDIF
\RETURN \texttt{tax\_annot}
\end{algorithmic}
\caption{pseudocode for taxonomic assignation through informative reads: \texttt{tax\_annot = informative(q,$\alpha$)} }
\label{alg:inforead}
\end{algorithm}

Then, procedure \texttt{tax\_annot = informative(q,$\alpha$)} has three possible outcomes:
\begin{itemize}
 \item \texttt{unknown} if the query is too far from the reference database
 \item \texttt{ambiguous} if there are several references in the neighborhood of $q$, but of different taxa
 \item \texttt{a taxon} if all the references in the neighborhood of $q$ belong to the same taxon
\end{itemize}

\paragraph{Step 4:} Programme \texttt{informative} is run as a loop over all reads $q \in \mathbf{Q}_i$. For $2.10^5$
queries, it takes less than 30 seconds in python on an standerd laptop with Linux Ubuntu. If there are $n$ queries, this gives a vector 
of strings $v$ of length $n$. 
Step 4 simply is to produce a table which counts the number $n_i$ of occurrences of item $i$ (where $i$ can be \texttt{unknown}, 
\texttt{ambiguous} or any taxon for which at least a read is informative). This step is called 
\texttt{diagno\_syst(tax\_annot)}.

\paragraph{Step 5:}Step 4 is run for any barconding gap $\alpha \in \{0,\theta\}$. Steps 3, 4 and 5 have been associated 
into a single program called \texttt{diagno\_syst\_loop}, and reads\\
\\
\begin{algorithm}[H]
\begin{algorithmic}[1]
\STATE \textbf{input:} $\mathbf{Q}$ 
\STATE \textbf{input}  $\theta$
\FOR{$\alpha \in 0,\ldots,\theta$}
\STATE $v \leftarrow [.]$
\FOR{$q \in \mathbf{Q}$}
\STATE $v[q] \leftarrow \texttt{informative}(q,\alpha)$
\ENDFOR
\STATE $x_{\alpha} \leftarrow \texttt{diagno\_syst}(v)$
\ENDFOR
\RETURN $\mathbf{X} = [x_0|\ldots|x_\theta]$
\end{algorithmic}
\caption{pseudocode for \texttt{diagno\_syst\_loop}}
\label{alg:diagnoloop}
\end{algorithm}

\paragraph{Notes:} First, We have not looked at an optimization of calculation time here, but have focused on exact calculation. For example,
\cite{Liu2009} have proposed CUDASW++, an optimization of SW algorithm for or CUDA-enabled GPUs. See for example \cite{Okada2015} for 
recent background. Second, 
we have filtered the set $\mathbf{Q}$ by keeping queries with a length $300 \leq \ell(q) \leq 315$, as we wish a high homology 
between the query and some references. About one half of the queries only have been kept with this filter for each environmental sample. Third,
in the discussion section, we will compare the inventories produced this way with inventories produced with Mothur \cite{Schloss2009}.

\subsection{Classification of mismatches}

Detecting and quantifying eventual problems was done by directly comparing the abundance of taxa in the molecular inventory and the 
optical inventory. Abundances were divided into low abundance and high abundance. 
Low abundance in the optical inventories was defined as $\leq  1 \%$ relative abundance of a taxon, high abundance was $\geq 1 \%$. In 
the molecular inventories, low abundance was defined as a low number of amplicon reads $ \leq 1000$, high abundance then was defined as
$\geq 1000$ reads. Comparison was then done between the abundance of a taxon in the optical inventory and the abundance of reads for each 
barcoding gap. Hence, each taxon encountered in at least one of the inventories has been allocated a category per inventory
\begin{itemize}
 \item $\emptyset$: absent
 \item $\ell$: low abundance
 \item $A$: high abundance.
\end{itemize}
For example, a taxon absent in optics and present in low abundance in molecular inventories will be tagged $(\emptyset,\ell)$.
In an ideal case, both the list of species and their abundance would match exactly between both methods (only $(\ell,\ell)$ or $(A,A)$. 
In practice, this is not the case.\\
\\ 
The mismatches have been studied, and allocated to the following possible causes (first code for optics, second for molecular):
\begin{itemize}
 \item $(\emptyset, \ell)$: 
 \begin{itemize}
  \item the optical inventory could be made down to genus only; hence, the taxon was not identified to species and could therefore 
  not be addressed to one of the species sequences of the molecular database: code $e$
  \item the species has been detected with large gap $(\geq 15)$, hence is likely a match on another nearby species: code $i$
  \item there exists a discrepancy between DNA based classification and morphological based classification: code $a$
  \item the species is rare, has been detected in molecular inventory, but not in optical: code $b$
 \end{itemize}
 \item $(\emptyset, A)$: All codes $a,e,i$ can explain such a discrepancy as well (as in $(\emptyset,\ell)$) when the reads assigned to 
 a given taxon are abundant too. This is however not the case for code $b$ (rare species) where the reads cannot be abundant. Hence, code $e$
 cannot be found in such a situation.
 \item $(\ell,A)$ or $(A,\ell)$: It is highly likely that there has been an amplification problem: code $g$
 \item $(\ell, \emptyset)$
 \begin{itemize}
  \item as a symmetry to $(\emptyset, \ell)$, there exists a discrepancy between DNA based classification and morphological based 
  classification: code $a$ 
  \item the barcode of corresponding species, as as well as of nearby species, is absent from reference database: code $c$
  \item the optics has probably lead to count dead frustules: code $f$ 
 \end{itemize}
 \item $(A, \emptyset)$: 
 \begin{itemize}
  \item as in the case $(\ell, \emptyset)$, situations described by code $a$ or $c$ are possible
 \item code $f$, however, is unlikely, 
 \end{itemize}
\end{itemize}

The key for possible misclassification causes is given in table \ref{tab:codes}.\\
\\
The classification of mismatches par sample according to these codes has been done with expertise of diatomists who have done the optical 
inventories.

\section{Results}

There are three results per set of sample (Léman lake and Swedish rivers):
\begin{itemize}
 \item the optical inventories
 \item the molecular based inventories
 \item the likely causes for mismatch between both
\end{itemize}
Optical inventories are given in file \texttt{Léman\_optics.txt} for Léman lake and \texttt{Sweden\_optics.txt} for Swedish rivers. An example 
of molecular based inventory with the number of informative reads per gap (see methods) is given in file \texttt{L6.txt}. Mismatch classification 
are given in tables \ref{tab:leman} for Léman lake and table \ref{tab:swedish} for Swedish rivers.

\section{Discussion}

\subsection{Molecular based inventories}\label{subsec:inventories}

\paragraph{Similarity based data retrieval:} The pipeline presented here for supervised clustering can be summarized as a two steps process for each
query: $(i)$ a search for neighbors in the reference database, $(ii)$ a post-processing of this set. The search can be 
processed either by BLAST, or here by exact local alignment scores. We have selected the latter because it is easily
parallelizable. The key point is on post-processing. The question of allocating a query to a reference, knowing a reference 
database and a distance, has been thoroughly studied in many areas under the name of similarity search, or similarity based data retrieval 
(see e.g. \cite{Pestov2000}). The model behind taxonomic assignment is the model behind barcoding: there exists a barcoding
gap $\alpha$ such that if two sequences $x,y$ are at a distance $d(x,y) < \alpha$, they belong to the same taxon, 
whereas if they are at a distance $d(x,y)> \alpha$, they belong to different taxa. We discuss here how to assign a name knowing
a gap (notion of informative read), and how to extend the procedure to a set of barcoding gaps (notion of sliding barcoding gap).

\paragraph{Notion of informative read:} The notion of informative read is at the basis of our assignment pipeline, and is
developed here. For  a 
given barcoding gap $\alpha$, the neighborhood of a query is
retrieved as the set of references at distance $d \leq \alpha$. If all belong to same taxon in reference database, this taxon 
is assigned to the query. If not, the annotation is called ambiguous. If the neighborhood is the empty set, the annotation
is unknown. Let us illustrate this on the example given in figure 1. Let us have two different taxa shown here with blue dots,
called here taxon 1 and taxon 2. 
Each taxon is a clique, as they have been derived by a mixture of morphological and molecular based basis. Let us have 
two queries, one green one, and one red one. There are edges between
each query and all references at distance less that the barcoding gap $\alpha$ whatever their taxon. Both edges from green 
query are towards
nodes in taxon 1. Hence the read is informative and the label taxon 1 can be assigned to it. On the contrary, one edge from red 
query is towards taxon 1 and one is towards taxon 2. Then, red query is not informative, and assignment is ambiguous.

\paragraph{Sliding barcoding gap:} The barcoding gap is a value such that two reads separated by a distance larger than the
gap are expected to belong to different taxa, and to the same taxon if their distance is smaller than the gap. The gap 
depends on the taxonomic level. It is commonly accepted that an homology of $97\%$ or more is relevant for assignation
at species level. As the marker used here is 312 bp long, this leads to a gap of 9 bp (either mismatch or unitary indel). 
However, it is commonly accepted as well that the barcoding gap may depend on the clade, and is not uniform. Hence, we have 
implemented the procedure deciding whether a read is informative, or not, and the taxonomic assignment if it is informative 
for a range of barcoding gaps between $\mbox{gap} = 0$ (the more stringent choice) to $\mbox{gap}=20$, as it is not expected to 
have
distances larger than 20 bp (less than $93\%$ of homology) within the same species. This procedure has been run for all reads. 
This means that we have produced an array with $N$ rows if the sample is made of $N$ reads, and 21 columns, one column for each
barcoding gap between 0 and 20. At row $i$ and column $\alpha$, we have either a taxon if read $i$ is informative for gap 
$\alpha$, or the character \emph{ambiguous} or \emph{unknown}. Then, for each gap $\alpha$ with $0 \leq \alpha \leq 20$, and
for each taxon $t$ in the reference database, we have computed the number $n(t,\alpha)$ of reads which are informative with taxon 
$t$ at barcoding gap $\alpha$. Interestingly, for a given taxon $t$, the curve plotting the number $n(t,\alpha)$ of informative 
reads in the sample hitting a given 
taxon as a function of the barcoding gap $\alpha$ is often, but not always, unimodal. This means that this number can increase or decrease when the gap 
decreases. It
is likely that it increases for small values because the neighborhood (the set of specimen at distance equal to or less
than the gap) increases and specimens in it belong to the same taxon, and decreases for larger values because for some reads the 
neighborhood becomes taxonomically heterogeneous when the distance increases. This phenomenon is driven by the shape of
the reference database, and is under study with tools from multivariate analysis or machine learning. An example of such a 
phenomenon for one read is given in figure 2. It can be shown that if a read is informative for a given species at a 
given gap, it cannot be informative for another species at a different gap. The advantage of such an assignment procedure is that it permits 
assignment without a selection of a unique barcoding gap: it is an adaptive procedure, without a model for barcoding gap (either
constant or clade dependent).

\subsection{Classification of mismatches}

\paragraph{Interpretation of the results:}

The fraction of perfect matches between morphological based and molecular based inventories is of 14 \% for Léman lake, and 7 \% for 
Swedish rivers. 
This could appear as desperately low. The main causes for mismatches are 
\begin{itemize}
 \item for Léman lake: case $b$ (rare species not recognized in optics) for 41 \%, then $c$ (absence of barcode in R-Syst) for 17 \%,
 $i$ (likely match on a nearby species) for 14 \%, and $a$ (mismatch between molecular based and morphological based taxonomy) for 11 \%.
\item for Swedish rivers: $a$ (mismatch between molecular based and morphological based taxonomy) for 26 \%, $c$ (absence of barcode in 
R-Syst) for 24 \%, then   $b$ (rare species not recognized in optics) for 13 \%, $i$   (likely match on a nearby species) for 9 \% and
$e$ (optical inventory at genus level only) for 8 \%
\end{itemize}
All others causes can be considered as negligible (less than 5 \%). It appears that the main sources of mismatches are not due to the
lack of resolution of molecular marker or accuracy of the pipeline, but to taxonomical and technical difficulties, which can be arranged 
in three 
categories: 
\begin{itemize}
 \item a still ongoing mismatch for some groups or clades between morphological based and molecular based taxonomy (case $a$)
 \item a difficulty in optics to accurately detect rare species (case $b$)
 \item the incompleteness of the reference data base (cases $c$ and $i$), preventing an accurate molecular based inventory.
\end{itemize}
Hence, the conditions for a molecular based inventory in metabarcoding to be accurate are that $(i)$ there is an agreement between 
morphological based and molecular based systematics and $(ii)$ the reference molecular database encompasses the whole diversity of
the sampled communities. If those conditions are fulfilled, there is a slight advantage for molecular based inventories which could
be more accurate for detecting rare species.

\paragraph{Incompleteness of the reference database:} Regarding the Swedish dataset, about 50\% of all taxa found in the optical inventory 
were not present in the reference database, so were impossible to find by the NGS method. Checking the different streams, we found that 
many of the diatom species dominating the streams of the boreal region are not present in database and that taxa from acid streams are 
especially missing. Examples of Swedens most frequent taxa missing from the reference database are \emph{Brachysira neoexilis} Lange-Bertalot, 
\emph{Eunotia incisa} W. Smith \& W. Gregory and \emph{Eunotia implicata} N\"orpel, Lange-Bertalot \& Alles. Additionally, taxa from high 
mountain regions are missing as well. Best represented are typical agricultural streams from the non-boreal region of Sweden, ecologically 
most similar to Central Europe. On the contrary, 17\% only
of taxa found optically in Léman lake are absent in the reference database. They are therefore impossible to detect in the molecular 
inventories. It is the case for taxa such as species of the \emph{Encyonopsis} genus or \emph{Calonei bacillum}, \emph{Navicula 
radiosafallax}, \emph{N. utermoehlii}, \emph{Nitzschia lacuum}. There are also several species of the \emph{Achnanthidium} genus 
(\emph{A. catenatum}, \emph{A. eutrophilum}) which were not detected in the molecular inventory but were observed in microscopy for 
the same reason.
This is in line with recent works of \cite{Abad2016} which ranked completeness/incompleteness of reference database as first among causes 
for mismatch between optical and molecular based inventories.

\paragraph{Discrepancy between molecular based and morphological based taxonomy:} Such a discrepancy is a major limit for building an 
agreement between 
classical optical inventories and metabarcoding. A most prominent example for this category of problems is the genus 
\emph{Fragilaria}, recorded as a problem in all but four of the 20 samples of the Swedish dataset. Even if a number of recent publications 
are trying to unravel the taxonomy of this genus, its identification and separation between species are far from clear. Identification 
literature gives often no clear limits between species, some features are only visible in electronic microscopy, and most of all, different 
references are often giving different characters for the different species (see e.g. the case of \emph{F.gracilis} {\O}strup, which is very 
common in Fennoscandia, but not so much in Central Europe, where it often is identified as 
\emph{F. rumpens} (K\"utzing) Lange-Bertalot instead). This leads to problems when 
trying to separate these species in optical inventories, and also especially when giving a name to a species in a reference database leading 
in turn to problems when comparing optical and molecular inventories to each other. In some circumstances, we can today not even say which 
of the given names in 
an inventory are correct, but have to go back and harmonize the way of identifying species. There is as well a need to make close studies of 
molecular and optical characters of taxa which 
were until now considered as a species, and study the accuracy of separations to closely related species. If none can be found, it may lead 
to pool species in order to enable 
clear identifications. Other taxon complexes suffer of similar problems, represented in the Swedish dataset by for example 
\emph{Achnanthidium minutissimum} and related species, \emph{Nitzschia palea} and related species, \emph{Navicula cryptocephala} and 
\emph{N. cryptotenella} and related species, \emph{Eolimnia minima} and related species, certain \emph{Gomphonema} species, the genera 
\emph{Staurosira}, \emph{Ulna}, \emph{Encyonema} \emph{Cymbella}, \emph{Mayamaea}, \emph{Amphora}, and \emph{Planothidium}. 

\subsection{Comparison with existing pipelines}

Mismatches between optical based and molecular based inventories can come from biases in one of the inventory (or both ...), 
as well as 
disagreements between morphology based and molecular based systematics. Hence, we have tried to minimize the possible biases due to the 
computing phase in building a molecular based inventory. Therefore, we have favored exact calculations, i.e. calculation of all 
local alignments between queries and references. We have compared our inventories with the ones issued on same datasets with standard tools, 
here Mothur \cite{Schloss2009}.

\paragraph{Comparison with Mothur: methods}
Mothur requires that the reference database is aligned, whereas diagno-syst does not. For each query, mothur searches the read in reference 
database closer to the query (the best hit), makes a local alignment between both, and provides some information on the quality of the alignment. 
When the 
quality of the alignment is considered as sufficient (some threshold have to be defined for that), next step is to accept the alignment as 
good, look at the identity of the aligned reference, look at its name, and transfer it to the query, considered as an element of the 
inventory. Different tools can be selected as options, as kmers, suffix tree or BLAST for searching closest reference, and global or local 
alignment for the
quality of the alignment. We have selected a search with kmers, and local alignment with gotoh. We have been stringent with the quality of 
the alignment (300 aligned bp or more among 312 bp). Beyond the necessity to have an aligned reference database, and use of heuristics by 
mothur, the main difference between mothur and diagno-syst is a taxonomic annotation from the best hit by mothur, and by an informative read 
by diagno-syst. Diagno-syst is more stringent, in the sense that, for a given gap, it requires that all neighbors in reference 
database at distance less than the gap belong to the same taxon. Then the name is transferred. It is therefore expected that Mothur has 
more false 
positive, and diagno-syst more false negative. 

\paragraph{Comparison with Mothur: an example} A comparison between both inventories has been made for each sample of swedish rivers and L\'eman 
lake, and 
results for one sample (UR\_775, Swedish rivers) are presented here as an example. The comparison is given in file \texttt{UR\_757\_compare.txt}. 
The sample is oligotrophic and neutral, and from the pristine 
mountains. 45 species have been recognized by optics, 69 with diagno-syst, and 134  with Mothur. 10 false positive have been produced by 
diagno-syst, and 34 by Mothur. It has been more difficult to quantify the amount of false-negatives, as into the category “taxonomical/barcoding 
problems”: they might have been hit but as a closely related taxon. Another possibility is that there is a mismatch between optical and molecular 
based species delineation, like for \emph{Cocconeis placentula}. Hence, we have counted the taxa that diagno-syst and Mothur found and were absent 
from optical inventory, and assessed which were probably (e.g. earlier found) or likely (i.e. oligotrophic, clean-water taxa) in the sample. This
represents 13 species (out of 24 false positive) for diagno-syst, and 25 (out of 89) for Mothur. This shows that diagno-syst 
inventories are closer to optical based inventories than mothur inventories. This is probably due to the fact that, in order to be quick, mothur 
runs one local alignment only, and diagno-syst, in order to be more exact, runs them all (in this early version). Mothur assignment relies on the 
alignment with the best hit only, whereas diagno-syst relies on the alignment with all reference reads in a given neighborhood, and hence is more
stringent. 

\paragraph{Comparison with Mothur at genus level:} This can be checked by a comparison between both approaches at the genus level, where taxonpic discrepancies between different 
approaches are less developed. The result is given in table \texttt{UR\_757\_compare\_genera.txt}. Looking at it indicates in nice global 
correlation (better seen on a log-log scale), but a few important discrepancies. A major one is for genus \emph{Cymbella}, and all the other ones 
are in a long tail of rare genera, found by Mothur and not by diagno-syst \emph{Placoneis, Craticula, Acanthoceras, Fallacia, ...}. This comparison 
between Mothur and diagno-syst has to be studied further and in more details.


\section{Conclusions}

We have designed and run a new pipeline, called \verb%diagno-syst% for molecular based inventories in metabarcoding. We have produced a pipeline 
which enables an industrialization of production of such inventories. It relies on computing exact distances, without heuristics, between each 
read of an environmental sample and each sequence of a reference database. Such distances are then used to build molecular based inventories 
on environmental samples. We have compared the results of inventories of freshwater benthic diatom communities with 
optical based inventories on two contrasted sets of environmental
samples: 10 samples from Léman lake, and 20 samples from Swedish rivers. All samples have been inventoried optically, and with metabarcoding,
with a same protocol, and using a reference database specifically designed for freshwater diatoms (R-Syst::diatoms, see \cite{Rimet2016}). We
have compared the outputs of our pipeline with the standard existing one for metabarcoding inventories in microbial communities: Mothur. We have 
found that our pipeline is more stringent: both recovered species which had been recognized optically, but they differed in the number and types 
of false positive: more positive, and sometimes taxonomically more distant from what could be expected, with Mothur. The higher quality and accuracy
of diagno-syst inventories is paid by a longer computation time, which is reasonable with parallelization (which has been implemented).\\
\\
We have studied a main 
question: is it possible to quantify the fraction of match/mismatch between both ways of performing an inventory on a same sample, split 
according to potential causes? We have been interested in answering to two further questions: first, a focus on rare species, by definition 
difficult to detect, but a key component of biodiversity, and, second, having two contrasted datasets handled with the same protocol permits 
to have an idea of the dependence of the result on regional idiosyncrasies.\\
\\
In both datasets, the largest causes of mismatches between molecular and optical dataset were $a$ - Discrepancy 
between DNA-data and classical taxonomy, $b$ - rare species, $c$ - no barcode in R-Syst and $i$ - species detected for gaps $\geq 15$. 
However, we found that the plausible main causes of discrepancies are different for datasets: In Léman lake, most mismatches were caused 
by taxa that were found by metabarcoding but not in microscope. This category was interpreted as rare species, which simply had been 
overlooked in 
the microscope, as only 400 valves are too be counted. As our knowledge of diatoms benthic communities in Léman lake is fairly good, 
it is possible to state in most of cases that those species actually 
have been found earlier in the frequent and long-term monitoring of this lake. So we can assert with high confidence that the NGS method is 
correct here, and that the NGS method is better in finding rare taxa. On the other 
hand, in the Swedish dataset most mismatches were caused by $(i)$ a discrepancy between DNA-based and classical taxonomy 
and $(ii)$ by optical taxa missing in the reference database.  Furthermore, a comparison between Mothur and diagno-syst showed that the long tail
of rare species or genera provided by Mothur is not fully trustable: it is probably over-estimated. Most of these discrepancies are due to an 
underrepresentation of boreal taxa in the reference database.\\
\\
As a consequence, much effort still has to be put into $(i)$ implementation of exact calculations (sequence comparisons) for comparing queries 
with a 
reference database in order to minimize an overestimation of the long tail of diversity, $(ii)$ diatom taxonomy, to unravel DNA based and optical 
based species delineation, and $(iii)$ completing reference databased with more species from underrepresented regions. 

\paragraph{Acknowledgements:} It is a pleasure to acknowledge a definite input of Yec'han Laizet (BioGeCo) for an earlier version of pipeline 
\texttt{diagno\_syst}, several discussions on metabarcoding of protists with D. Debroas, the earlier work of and discussions with Lena\"ig 
Kermarrec 
as PhD student on metabarcoding in Carrtel with tight collaborations with BioGeCo. This work has been supported by ONEMA project 
"Developpement d’outils de bio-indication phytobenthos et 
macroinvert\'ebr\'es benthiques pour les cours d’eau de Mayotte" given to AB, project e-Biothon given to AF, DARI project "biodiversiton" 
i2015037360 in 2015 and 2016 for access to National Computing center IDRIS, given to AF, and a one year CDD given by CEBA 
(Centre d'\'etude de la biodiversité amazonienne) ANR Project to AF, which supported an earlier version of diagno-syst. The Swedish contribution 
to this study was jointly funded by The Swedish Agency for Marine and Water Management, SwAM, and the Environmental Monitoring and Assessment 
(Fortlöpande miljöanalys (Foma) och miljöövervakning) research program at the Swedish University of Agricultural Sciences.

\clearpage
\section{Tables}

\begin{table}[h]
\begin{center}
\begin{tabular}{l|l}
\hline
code & meaning \\
\hline
a & Discrepancy between DNA-data and classical taxonomy	\\
b & rare species \\	
c &  no barcode in R-Syst \\	
e &  Genus level in optics only	\\
f &  If barcode in R-Syst: no DNA, dead frustule \\	
g &  amplification problem \\	
h &  other reasons \\	
i &  species detected for gaps $\geq$ 15 	\\
k & ok \\
\hline
\end{tabular}
\caption{Coding of the possible reasons for discrepancy between optical based and molecular based inventories\label{tab:codes}}
\end{center}
\end{table}


\vspace*{1cm}

\begin{table}
\begin{center}
\begin{tabular}{l|rrrrrrrrr}
\hline
 & a  & b   & c   & e   & f & g & h & i & k \\  
\hline	   
L1 & 12 & 27 & 11 & 0 & 3 & 0 & 0 & 5 & 10 \\
L2 & 7 & 25 & 12 & 0 & 1 & 0 & 0 & 6 & 11  \\
L3 & 12 & 26 & 9 & 0 & 1 & 0 & 0 & 5 & 8  \\
L4 & 6 & 38 & 11 & 0 & 2 & 0 & 0 & 9 & 8  \\
L5 & 7 & 27 & 10 & 0 & 1 & 0 & 0 & 10 & 8  \\
L6 & 7 & 27 & 12 & 0 & 1 & 0 & 0 & 12 & 9  \\
L7 & 6 & 27 & 11 & 0 & 3 & 0 & 0 & 11 & 11  \\
L8 & 7 & 25 & 8	 & 0 & 2 & 0 & 0 & 10 & 6  \\
L9 & 5 & 18 & 16 & 0 & 4 & 0 & 1 & 9 & 11  \\
L10 & 8 & 33 & 15 & 0 & 2 & 0 & 1 & 15 & 10  \\
\hline
Sum & 77 & 273 & 115 & 0 & 20 & 0 & 2 & 92 & 92  \\										
(\%, rounded) & 11 & 41 & 17 & 0 & 3 & 0 & 0 & 14 & 14  \\
\hline
\end{tabular}
\end{center}
\caption{Counting the number of species per sample (in row) and per possible cause of discrepency (in columns) for 10 samples of
Léman lake. See table \ref{tab:codes} for the meaning of the codes.\label{tab:leman}}
\end{table}


\clearpage
\begin{table}
\begin{center}
\begin{tabular}{l|rrrrrrrrr}
\hline
	   & a  & b   & c   & e   & f & g & h & i & k \\  
\hline	   
UR\_787	   & 20 & 5   & 9   & 3   & 0 & 2 & 4 & 7 & 3 \\
UR\_36	   & 32 & 15  &	9   & 9   & 0 & 0 & 7 & 8 & 3 \\
UR\_775	   & 24 & 11  & 25  & 5   & 0 & 2 & 3 & 4 & 3 \\
UR\_1	   & 30 & 12  &  20 & 7   & 0 & 1 & 6 & 7 & 6 \\
UR\_38	   & 29 & 9   &	31  & 8   & 0 & 5 & 4 & 8 & 2 \\
UR\_789	   & 30 & 10  &	28  & 5   & 0 & 3 & 3 & 6 & 1 \\
UR\_SAP45A & 7	& 5   &  27 & 5   & 0 & 2 & 1 & 3 & 1 \\
UR\_764	   & 6	& 2   & 14  & 3   & 0 & 0 & 4 & 4 & 5 \\
UR\_39	   & 29	& 17  & 25  & 4   & 0 & 1 & 5 & 7 & 11 \\
UR\_790	   & 30	& 21  & 30  & 6   & 0 & 1 & 4 & 10 & 13 \\
UR\_803	   & 23	& 1   & 8   & 3   & 2 & 1 & 4 & 11 & 2 \\
UR\_27	   & 35	& 11  & 16  & 6   & 0 & 0 & 3 & 4 & 4 \\
UR\_771	   & 6	& 3   & 12  & 3   & 0 & 0 & 3 & 1 & 2 \\
UR\_785    & 25	& 20  & 23  & 5   & 0 & 4 & 4 & 9 & 5 \\
UR\_SAM36A & 8	& 6   & 6   & 5   & 1 & 6 & 0 & 2 & 5 \\
UR\_756	   & 6	& 2   & 7   & 2   & 0 & 2 & 2 & 3 & 1 \\
UR\_53     & 4	& 5   & 8   & 4   & 0 & 1 & 3 & 5 & 2 \\
UR\_766	   & 26	& 12  & 4   & 8   & 1 & 3 & 2 & 1 & 9 \\
UR\_26	   & 34	& 14  & 17  & 11  & 1 & 2 & 6 & 21 & 14 \\
UR\_757	  & 29	& 5   & 17  & 9   & 2 & 7 & 2 & 12 & 4 \\
\hline									
sum 	  	& 433 & 186 & 336 & 111 & 7 & 43 & 70 & 133 & 96 \\
(\%, rounded)   & 30  & 13  & 24  & 8   & 0 &  3 & 5  & 9 & 7 \\
\hline 
\end{tabular}
\end{center}
\caption{Counting the number of species per sample (in row) and per possible cause of discrepancy (in columns) for 20 samples of
swedish rivers. See table \ref{tab:codes} for the meaning of the codes.\label{tab:swedish}}
\end{table}

\clearpage
\section{Figures}

\begin{tikzpicture}
 \coordinate (A) at (0,0) ;
 \coordinate (B) at (0.5,1.3) ;C
 \coordinate (C) at (1.7, 0.5) ;
 \coordinate (D) at (1.5, 2.3) ;
 \coordinate (E) at (3.0, 1.6) ;
 \coordinate (T1) at (1.5 , 3.5) ;
 \fill[color=blue] (A) circle (0.1);
 \fill[color=blue] (B) circle (0.1);
 \fill[color=blue] (C) circle (0.1);
 \fill[color=blue] (D) circle (0.1);
 \fill[color=blue] (E) circle (0.1);
 \draw[color=blue] (A) -- (B);
 \draw[color=blue] (A) -- (C);
 \draw[color=blue] (B) -- (C);
 \draw[color=blue] (B) -- (D);
 \draw[color=blue] (B) -- (E); 
 \draw[color=blue] (C) -- (D);
 \draw[color=blue] (C) -- (E);
 \draw[color=blue] (D) -- (E);
 \draw (T1) node {Taxon 1};
 \coordinate (A2) at (5.0,0) ;
 \coordinate (B2) at (5.5, 1.3) ;
 \coordinate (C2) at (6.4, 0.7) ;
 \coordinate (D2) at (5.2, 2.3) ;
 \coordinate (T2) at (5.5 , 3.5) ;
 \fill[color=blue] (A2) circle (0.1);
 \fill[color=blue] (B2) circle (0.1);
 \fill[color=blue] (C2) circle (0.1);
 \fill[color=blue] (D2) circle (0.1);
 \draw[color=blue] (A2) -- (B2);
 \draw[color=blue] (B2) -- (C2);
 \draw[color=blue] (B2) -- (D2);
 \draw[color=blue] (A2) -- (C2); 
 \draw[color=blue] (A2) -- (D2);
 \draw[color=blue] (C2) -- (D2); 
\draw (T2) node {Taxon 2}; 
\coordinate (Q1) at (2.7, 2.8)	;
\fill[color=green] (Q1) circle (0.1) ; 
\draw[color=green] (Q1) -- (D) ;
\draw[color=green] (Q1) -- (E) ;
\coordinate (Q2) at (4.2, 1.7)	;
\fill[color=red] (Q2) circle (0.1); 
\draw[color=red, dashed] (Q2) -- (E) ;
\draw[color=red, dashed] (Q2) -- (B2) ;
\end{tikzpicture}

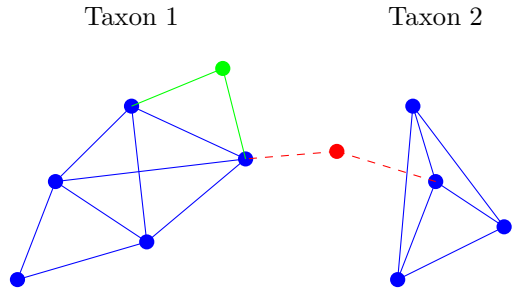
\paragraph{Figure 1:} Blue dots are references. Green and red dots represent two queries. References have been categorized
by a mixture of molecular and morphological based assignment. Hence they are depicted as (nearly) cliques. There is an edge 
between a 
query and a reference when the distance 
between them is less than $\alpha$. Green query is an informative read, because both edges lead to the same taxon, 
whereas red query is ambiguous, as both edges lead to two different taxa. See text for details.

\vspace*{2cm}

\begin{tikzpicture}
 
 \coordinate (Q) at (0,0) ;
 \coordinate (A1) at (2.5,3.3) ;
 \coordinate (A2) at (2.1, 1.5) ;
 \coordinate (A3) at (1.5, 2.3) ;
 \coordinate (A4) at (3.0, 2.6) ;
 \coordinate (A5) at (1.5 , 3.5) ;
 \coordinate (B1) at (3.5, 4.2) ;
 \coordinate (B2) at (4.2, 4.6) ;  
 \coordinate (B3) at (5.2, 4.4) ;   
 \fill[color=red] (Q) circle (0.1); 
 \fill[color=blue] (A1) circle (0.1);
 \fill[color=blue] (A2) circle (0.1);
 \fill[color=blue] (A3) circle (0.1);
 \fill[color=blue] (A4) circle (0.1);
 \fill[color=blue] (A5) circle (0.1);
 \fill[color=green] (B1) circle (0.1);
 \fill[color=green] (B2) circle (0.1); 
 \fill[color=green] (B3) circle (0.1);  
 \draw (2.0,0) arc (0:90:2.0) ;
 \draw (0.8, 1.1) node {$n=0$};
 \draw (0.8, 0.7) node {(unknown)};
 \draw (4.5,0) arc (0:90:4.5) ;
 \draw (0.8, 3.0) node {$1 \leq n \leq 5$};
 \draw (1.9, 4.8) node {$n>5$ (ambiguous)};

\end{tikzpicture}

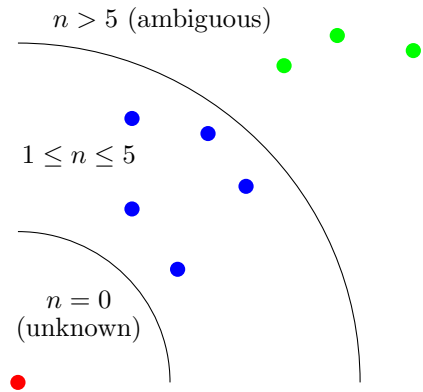
\paragraph{Figure 2:} The red dot is a read. The blue dots and green dots are two species in the reference database, 
respectively. Black quarter of circle represent areas at distance equal to the radius of the circle. For small distances,
the query has no it $(n=0)$. The status is \emph{unknown}. For some region, the read can have between 1 and 5 hits of references
belonging all to species blue $(1 \leq n \leq 5)$.The status is informative for species blue. Beyond a given radius, the read has 
more than 5 hits $(n> 5)$, but on references belonging to either blue or green species. The status is \emph{ambiguous}.

\clearpage


\begin{thebibliography}{99}

\bibitem{Abad2016}
D.~Abad, A.~Albaina, M.~Aguirre, A.~Laza-Martínez, I.~Uriarte, A.~Iriarte,
  F.~Villate, and A.~Estonba.
\newblock {Is metabarcoding suitable for estuarine plankton monitoring? A
  comparative study with microscopy}.
\newblock {\em Mar. Biol.}, 163:149--161, 2016.

\bibitem{AFNOR2014}
AFNOR.
\newblock {NF EN 14407 - Qualité de l'eau - Guide pour l'identification et le
  dénombrement des échantillons de diatomées benthiques de rivières et de
  lacs.}
\newblock {\em AFNOR}, pages 1--13, 2014.

\bibitem{AFNOR2016}
AFNOR.
\newblock {NF T90 354 - Qualité de l'eau - Échantillonnage, traitement et
  analyse de diatomées benthiques en cours d'eau et canaux.}
\newblock {\em AFNOR}, pages 1--79, 2016.

\bibitem{Altschul1990}
S.~F. Altschul, W.~Gish, W.~Miller, E.~W. Myers, and D.~J. Lipman.
\newblock Basic local alignment search tool.
\newblock {\em Journal of Molecular Biology}, 215:403--410, 1990.

\bibitem{Bachy2013}
C.~Bachy, J.~R. Dolan, P.~L\'opez-Garcia, P.~Deschamps, and D.~Moreira.
\newblock Accuracy of protist diversity assessments: morphology compared with
  cloning and direct pyrosequencing of 18{S} r{RNA} genes and {ITS} regions
  using the conspicuous tintinnid ciliates as a case study.
\newblock {\em ISME}, 7:244--255, 2013.

\bibitem{Bik2012}
H.~M. Bik, D.~L. Porazinska, S.~Creer, J.~G. Caporaso, R.~Knight, and W.~K.
  Thomas.
\newblock Sequencing our way towards understanding global eukaryotic
  biodiversity.
\newblock {\em Trends in Ecology and Evolution,}, 27:233–243, 2012.

\bibitem{Bruder2007}
K.~Bruder and L.~K. Medlin.
\newblock {Molecular assessment of phylogenetic relationships in selected
  species/genera in the naviculoid diatoms (Bacillariophyta). I. The genus
  Placoneis}.
\newblock {\em Nova Hedwigia}, 85:331–352, 2007.

\bibitem{Bucklin2016}
A.~Bucklin, P.~Lindeque, N.~Rodriguez-Ezpelata, A.~Albaina, and M.~Lehtiniemi.
\newblock Metabarcoding of marine zooplankton: prospects, progress and
  pitfalls.
\newblock {\em J. Plankton Res.}, 38(3):393--400, 2016.

\bibitem{Clarke2014}
L.~J. Clarke, J.~Soubrier, L.~S. Weyrich, and A.~Cooper.
\newblock Environmental metabarcodes for insects: in silico {PCR} reveals
  potential for taxonomic bias.
\newblock {\em Molecular Ecology}, 14:1160–1170, 2014.

\bibitem{Cowart2015}
D.~A. Cowart, M.~Pinheiro, O.~Mouchel, M.~Maguer, J.~Grall, J.~Min\'e, and
  S.~Arnaud-Haond.
\newblock Metabarcoding is powerful yet still blind: A comparative analysis of
  morphological and molecular surveys of seagrass communities.
\newblock {\em PloS One}, DOI:10.1371/journal.pone.0117562:1--26, 2015.

\bibitem{Dafforn2014}
K.~A. Dafforn, D.~J. Baird, A.~A. Chariton, M.~Y. Sun, M.~V. Brown, S.~L.
  Simpson, B.~P. Kelaher, and E.~L. Johnston.
\newblock Faster, higher and stronger? the pros and cons of molecular faunal
  data for assessing ecosystem condition.
\newblock {\em Advances in Ecological Research}, 51:1--40, 2014.

\bibitem{Debroas2015}
D.~Debroas, M.~Hugoni, and I.~Domaizon.
\newblock Evidence for an active rare biosphere within freshwater protists
  community.
\newblock {\em Molecular Ecology}, 24(6):1236–1247, 2015.

\bibitem{Guillou2013}
L.~Guillou, D.~Bachar, S.~Audic, D.~Bass, C.~Berney, L.~Bittner, C.~Boutte,
  G.~Burgaud, C.~de~Vargas, J.~Decelle, J.~del Campo, J.~R. Dolan, M.~Dunthorn,
  B.~Edvardsen, M.~Holzmann, W.~H. C.~F. Kooistra, E.~Lara, N.~Le~Bescot,
  R.~Logares, R.~Mahé, F. abnd~Massana, M.~Montresor, R.~Morard, F.~Not,
  J.~Pawlowski, I.~Probert, A.-L. Sauvadet, R.~Siano, T.~Stoeck, D.~Vaulot,
  P.~Zimmermann, and R.~Christen.
\newblock {The Protist Ribosomal Reference database (PR$^2$): a catalog of
  unicellular eukaryote Small Sub-Unit rRNA sequences with curated taxonomy}.
\newblock {\em Nucleic Acids Research}, 41(Database issue):D597–D604, 2013.

\bibitem{gusfield97}
D.~Gusfield.
\newblock {\em Algorithms on strings, trees, and sequences}.
\newblock Cambridge University Press, Cambridge, UK, 1997.

\bibitem{Hajibabaei2011}
M.~Hajibabaei, S.~Shokralla, X.~Zhou, G.~A.~C. Singer, and D.~J. Baird.
\newblock Environmental barcoding: a next generation sequencing approach for
  biomnitoring applications using river benthos.
\newblock {\em PLoS One}, 6(4):e17497, 2011.

\bibitem{hebert03}
P.~D.~N. Hebert, A.~Cywinska, S.~L. Ball, and J.~R. deWaard.
\newblock Biological identifications through {DNA} barcodes.
\newblock {\em Proc. R. Soc. Lond. B}, 270:313--321, 2003.

\bibitem{Hollingsworth2009}
P.~M. Hollingsworth, L.L. Forrest, J.~L. Spouge, and Hajibabaei~\& al.
\newblock A {DNA} barcode for land plants.
\newblock {\em PNAS}, 106:12794--12797, 2009.

\bibitem{Ji2013}
Y.~Ji, L.~Ashton, S.~M. Pedley, D.~P. Edwards, Y.~Tang, A.~Nakamura,
  R.~Kitching, P.~M. Dolman, P.~Woodcock, F.~A. Edwards, T.~H. Larsen, W.~X.
  Hsu, S.~Benedick, K.~C. Hamer, D.~S. Wilcove, C.~Bruce, X.~Wang, T.~Levi,
  M.~Lott, B.~C. Emerson, and D.~W. Yu.
\newblock Reliable, verifiable and efficient monitoring of biodiversity via
  metabarcoding.
\newblock {\em Ecology Letters}, 16:1245--1257, 2013.

\bibitem{Joly2014}
S.~Joly, T.~J. Davies, A.~Archambault, A.~Bruneau, A.~Derry, S.~W. Kembel,
  Peres-Neto P., J.~Vamosi, and T.~A. Wheeler.
\newblock Ecology in the age of {DNA} barcoding: the resource, the promise and
  the challenges ahead.
\newblock {\em Molecular Ecology}, 14:221--232, 2014.

\bibitem{Kermarrec2014}
L.~Kermarrec, A.~Franc, F.~Rimet, P.~Chaumeil, J.-M. Frigerio, J.-F. Humbert,
  and A.~Bouchez.
\newblock A next-generation sequencing approach to river biomonitoring using
  benthic diatoms.
\newblock {\em Freshwater Science}, 33:349--363, 2014.

\bibitem{Kermarrec2013}
L.~Kermarrec, A.~Franc, F.~Rimet, P.~Chaumeil, J.-F. Humbert, and A.~Bouchez.
\newblock Next-generation sequencing to inventory taxonomic diversity in
  eukaryotic communities: a test for freshwater diatoms.
\newblock {\em Molecular Ecology Resources}, 13:607--619, 2013.

\bibitem{Kunin2010}
V.~Kunin, A.~Engelbrektson, H.~Ochman, and P.~Hugenholtz.
\newblock Wrinkles in the rare biosphere: pyrosequencing errors can lead to
  artificial inflation of diversity estimates.
\newblock {\em Environmental Microbiology}, 12(1):118–123, 2010.

\bibitem{Lindeque2013}
P.~K. Lindeque, H.~E. Parry, R.~A. Harmer, P.~J. Somerfield, and A.~Atkinson.
\newblock Next generation sequencing reveals the hidden diversity of
  zooplankton assemblages.
\newblock {\em PLoS One}, 8(11)(e81327.):doi:10.1371/journal.pone.0081327,
  2013.

\bibitem{Liu2009}
Y.~Liu, D.~L. Maskell, and B.~Schmidt.
\newblock {CUDASW++: optimizing Smith-Waterman sequence database searches for
  CUDA-enabled graphics processing units}.
\newblock {\em BMC Research Notes}, 2:73 doi: 10.1186/1756--0500--2--73, 2009.

\bibitem{Lopez-Garcia2003}
P.~L\'opez-Garc\'ia, H.~Philippe, F.~Gail, and D.~Moreira.
\newblock Autochthonous eukaryotic diversity in hydrothermal sediment and
  experimental microcolonizers at the mid-atlantic ridge.
\newblock {\em PNAS}, 100:697--702, 2003.

\bibitem{Lopez-Garcia2001}
P.~L\'opez-Garc\'ia, F.~Rodriguez-Valera, C.~Pedros-Alio, and D.~Moreira.
\newblock Unexpected diversity of small eukaryotes in deep-sea antarctic
  plankton.
\newblock {\em Nature}, 409:603--607, 2001.

\bibitem{Medinger2010}
R.~Medinger, V.~Nolte, R.~V. Pandey, S.~Jost, B.~Ottenwalder, Schlotterer C,
  and J.~Boenigk.
\newblock Diversity in a hidden world: potential and limitation of
  next-generation sequencing for surveys of molecular diversity of eukaryotic
  microorganisms.
\newblock {\em Molecular Ecology}, 19(Suppl. 1):32–40, 2010.

\bibitem{Moon2001}
S.~Y. Moon van~der Staay, R.~De~Wachter, and D.~Vaulot.
\newblock {Oceanic 18S rDNA sequences from picoplankton reveal unsuspected
  eukaryotic diversity}.
\newblock {\em Nature}, 409:607--610, 2001.

\bibitem{NW70}
S.~B. Needleman and C.~D. Wunsch.
\newblock A general method applicable to search for similarities in the
  amino-acid sequence of two proteins.
\newblock {\em J. Mol. Biol.}, 48:443--453, 1970.

\bibitem{Okada2015}
D.~Okada, F.~Ino, and K.~Hagihara.
\newblock {Accelerating the Smith-Waterman algorithm with interpair pruning and
  band optimization for the all-pairs comparison of base sequences}.
\newblock {\em BMC Bioinformatics}, 16:321 DOI 10.1186/s12859--015--0744--4,
  2015.

\bibitem{Pawlowski2012}
J.~Pawlowski, S.~Audic, S.~Adl, and \emph{et al.}
\newblock {CBOL} {P}rotist {W}orking {G}roup: Barcoding eukaryotic richness
  beyond the animal, plant, and fungal kingdoms.
\newblock {\em PloS Biology}, 10(11):e1001419., 2012.

\bibitem{Pawlowski2014}
J.~Pawlowski, F.~Lejzerowicz, and P.~Eslin.
\newblock Next-generation environmental diversity surveys of foraminifera:
  Preparing the future.
\newblock {\em Biol. Bull.}, 227:93--106, 2014.

\bibitem{Pedros-Alio2012}
Pedr\'os-Ali\'o.
\newblock The rare bacterial biosphere.
\newblock {\em Annual Rreview of Marine Science}, 4:449--466, 2012.

\bibitem{Pedros-Alio2007}
C.~Pedr\'os-Ali\'o.
\newblock Dipping into the rare biosphere.
\newblock {\em Science}, 315:192--193, 2007.

\bibitem{Pestov2000}
V.~Pestov.
\newblock On the geometry of similarity search: dimensionality curse and
  concentration of measure.
\newblock {\em Information Processing Letters}, 73:47--51, 2000.

\bibitem{Porazinska2009}
D.~Porazinska, R.~M. Giblin-Davis, L.~Faller, W.~Farmerie, N.~Kanzaki,
  K.~Morris, T.~O. Powers, A.~E. Tucker, W.~Sung, and W.~K. Thomas.
\newblock Evaluating high-throughput sequencing as a method for metagenomic
  analysis of nematode diversity.
\newblock {\em Molecular Ecology}, 9:1439–1450, 2009.

\bibitem{Reeder2009}
J.~Reeder and R.~Knight.
\newblock The "rare biosphere": a reality check.
\newblock {\em Nature methods}, 6(9):636--637, 2009.

\bibitem{Rimet2016}
F.~Rimet, P.~Chaumeil, F.~Keck, L.~Kermarrec, V.~Vasselon, M.~Kahlert,
  A.~Franc, and A.~Bouchez.
\newblock {R-Syst::diatom}: An open-access and curated barcode database for
  diatoms and freshwater monitoring.
\newblock {\em Database: The Journal of Biological Databases and Curation.},
  pages 1--21, doi:10.1093/database/baw016, 2016.

\bibitem{Schloss2009}
P.~D. Schloss, S.~L. Westcott, T.~Ryabin, J.~R. Hall, M.~Hartmann, and E.~B. et
  al.~(2009) Hollister.
\newblock Introducing mothur: Open-source, platform-independent,
  community-supported software for describing and comparing microbial
  communities.
\newblock {\em Appl. Environ. Microbiol.}, 75:7537–7541, 2009.

\bibitem{SW81}
P.~D. Smith and M.~S. Waterman.
\newblock Identification of common molecular subsequences.
\newblock {\em J. Mol. Biol.}, 147:195--197, 1981.

\bibitem{Sogin2006}
M.~L. Sogin, H.~G. Morrison, J.~A. Huber, D.~M. Welch, S.~M. Huse, P.~R. Neal,
  J.~M. Arrieta, and G.~J. Herndl.
\newblock Microbial diversity in the deep sea and the underexplored ‘‘rare
  biosphere’’.
\newblock {\em PNAS}, 103(32):12115--12120, 2006.

\bibitem{Stoof-Leichsenring2012}
K.~R. Stoof-Leichsenring, L.~S. Epp, M.~H. Trauth, and R.~Tiedmann.
\newblock Hidden diversity in diatoms of kenyan lake naivasha: a genetic
  approach detects temporal variation.
\newblock {\em Molecular Ecology}, 21:1918–1930, 2012.

\bibitem{Taberlet2012}
P.~Taberlet, E.~Coissac, M.~Hajibabaei, and L.~Rieseberg.
\newblock Environmental {DNA}.
\newblock {\em Molecular Ecology}, 2:1789--1793., 2012.

\bibitem{Tanabe2016}
A.~S. Tanabe, S.~Nagai, K.~Hida, M.~Yasuike, A.~Fujiwara, Y.~Nakamura,
  Y.~Takano, and S.~Katakura.
\newblock {Comparative study of the validity of three regions of the 18S-rRNA
  gene for massively parallel sequencing-based monitoring of the planktonic
  eukaryote community.}
\newblock {\em Molecular Ecology Resources}, 16:402--414, 2016.

\bibitem{APGIII}
{The Angiosperm Phylogeny Group}.
\newblock {An update of the Angiosperm Phylogeny Group classification for the
  orders and families of flowering plants: APG III}.
\newblock {\em Botanical Journal of the Linnean Society}, 161(2):105--121 DOI
  10.1111/j.1095--8339.2009.00996.x, 2009.

\bibitem{Zimmerman2015}
J.~Zimmerman, G.~Gl\"ockner, R.~Jahn, N.~Enke, and B.~Gemeinholzer.
\newblock Metabarcoding vs. morphological identification to assess diatom
  diversity in environmental studies.
\newblock {\em Molecular Ecology}, 15:526–542, 2015.

\bibitem{Zinger2012}
L.~Zinger, A.~Gobet, and T.~Pommier.
\newblock Two decades of describing the unseen majority of aquatic microbial
  diversity.
\newblock {\em Molecular Ecology}, 21:1878–1896, 2012.

\end{thebibliography}

\end{document}